# Covered Clause Elimination

Marijn Heule*  Matti Järvisalo†  Armin Biere
TU Delft, The Netherlands  Univ. Helsinki, Finland  JKU Linz, Austria

**Abstract**

Generalizing the novel clause elimination procedures developed in [1], we introduce *explicit* (CCE), *hidden* (HCCE), and *asymmetric* (ACCE) variants of a procedure that eliminates *covered clauses* from CNF formulas. We show that these procedures are more effective in reducing CNF formulas than the respective variants of blocked clause elimination, and may hence be interesting as new preprocessing/simplification techniques for SAT solving.

## 1 Introduction

Simplification techniques applied both before (i.e., in preprocessing) and during search have proven integral in enabling efficient conjunctive normal form (CNF) level Boolean satisfiability (SAT)[1] solving for real-world application domains. Further, while many SAT solvers rely mainly on Boolean constraint propagation (i.e., unit propagation) during search, it is possible to improve solving efficiency by applying additional simplification techniques also during search. Noticeably, when scheduling *combinations* of simplification techniques during search, even quite simply ideas can bring additional gains by enabling further simplifications by other techniques.

Generalizing the clause elimination procedures developed in [1], in this paper we introduce *explicit* (CCE), *hidden* (HCCE), and *asymmetric* (ACCE) variants of a clause elimination procedure that eliminates what we call *covered clauses* from CNF formulas. We compare these procedures to the analogous variants BCE, HBCE, and ABCE (see Sect. 1.1) of blocked clause elimination [1, 2] w.r.t. *relative effectiveness*.

**Definition 1.** *Assume two clause elimination procedures $S_1$ and $S_2$ that take as input an arbitrary CNF formula F and each outputs a CNF formula that consists of a subset of F that is satisfiability-equivalent to F. Procedure $S_1$ is at least as effective as $S_2$ if, for any F and any output $S_1(F)$ and $S_2(F)$ of $S_1$ and $S_2$ on input F, respectively, we have that $S_1(F) \subseteq S_2(F)$; $S_2$ is not as effective as $S_1$ if there is an F for which there are outputs $S_1(F)$ and $S_2(F)$ of $S_1$ and $S_2$, respectively, such that $S_1(F) \subset S_2(F)$; and $S_1$ is more effective than $S_2$ if (i) $S_1$ is at least as effective as $S_2$, and (ii) $S_2$ is not as effective as $S_1$.*

This definition of relative effectiveness takes into account *non-confluent* elimination procedures, i.e., procedures that do not generally have a unique fixpoint and that may thus have more than one possible output for a given input. In fact, we show that out of the three covered clause elimination procedures, the explicit variant CCE is confluent. Extending the relative effectiveness hierarchy presented in [1] (see the solid arrows in Fig. 1), we show that the variants of covered clause elimination are more effective than their counterparts based on blocked clauses (see the dashed arrows in Fig. 1). In this sense, the elimination procedures introduced in this paper are proper generalizations of the techniques analyzed in [1]. This is interesting since it has been recently shown in [2] that already BCE is surprisingly effective, as it can—purely on the CNF level—implicitly perform a combination of structure-based circuit-level techniques, including the polarity-based Plaisted-Greenbaum CNF encoding and difference circuit simplifications. Here, the most effective technique is the asymmetric variant of covered clause elimination.

---

*Supported by Dutch Organization for Scientific Research under grant 617.023.611.
†Supported by Academy of Finland under grant #132812.

[1]We assume that the reader is familiar with basic concepts related to CNF satisfiability. When convenient we view a clause as a set of literals and a CNF formula as a set of clauses.

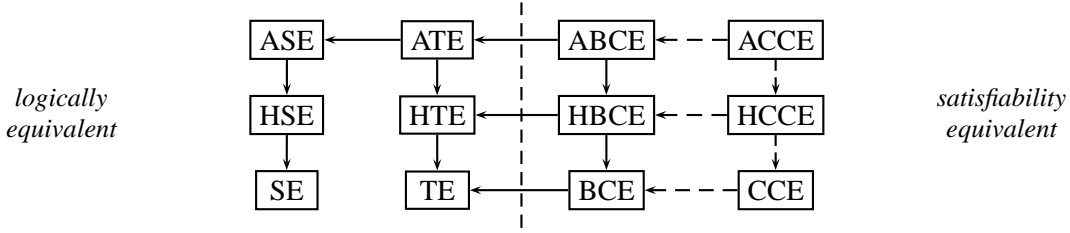

Figure 1: Relative effectiveness hierarchy of clause elimination procedures. An edge from $X$ to $Y$ means that $X$ is more effective than $Y$. A missing edge from $X$ to $Y$ means that $X$ is not as effective as $Y$. Transitive edges are missing from the figure for clarity. The dashed arrows present results of this paper.

## 1.1 Variants of Blocked Clause Elimination

The resolution rule states that, given two clauses $C_1 = \{l, a_1, \ldots, a_n\}$ and $C_2 = \{\bar{l}, b_2, \ldots, b_m\}$, the implied clause $C = \{a_1, \ldots, a_n, b_1, \ldots, b_m\}$, called the *resolvent* of $C_1$ and $C_2$, can be inferred by *resolving* on the literal $l$, and write $C = C_1 \otimes_l C_2$.

We compare the clause elimination procedures based on covered clauses to the following procedures [1] that eliminate blocked clauses [3]. Notice that out of these three, only BCE is confluent [1].

**BCE** Given a CNF formula $F$, a clause $C$ and a literal $l \in C$, the literal $l$ *blocks* $C$ w.r.t. $F$ if (i) for each clause $C' \in F$ with $\bar{l} \in C'$, $C \otimes_l C'$ is a tautology, or (ii) $\bar{l} \in C$, i.e., $C$ is itself a tautology[2]. Given a CNF formula $F$, a clause $C$ is *blocked* w.r.t. $F$ if there is a literal that blocks $C$ w.r.t. $F$. Removal of blocked clauses preserves satisfiability [3]. For a CNF formula $F$, *blocked clause elimination* (BCE) repeats the following until fixpoint: if there is a blocked clause $C \in F$ w.r.t. $F$, let $F := F \setminus \{C\}$. The CNF formula resulting from applying BCE on $F$ is denoted by BCE$(F)$.

**HBCE** Given a CNF formula $F$, we denote by $F_2$ the set of binary clauses contained in $F$. For a given clause $C \in F$, we denote by (*hidden literal addition*) HLA$(F,C)$ the *unique* clause resulting from repeating the following clause extension steps until fixpoint: if there is a literal $l_0 \in C$ such that there is a clause $(l_0 \vee l) \in F_2 \setminus \{C\}$ for some literal $l$, let $C := C \cup \{\bar{l}\}$. For a CNF formula $F$, a clause $C \in F$ is called *hidden blocked* if HLA$(F,C)$ is blocked w.r.t. $F$. *Hidden blocked clause elimination* (HBCE) repeats the following until fixpoint: if there is a hidden blocked clause $C \in F$, remove $C$ from $F$.

**ABCE** For a clause $C$ and a CNF formula $F$, (*asymmetric literal addition*) ALA$(F,C)$ denotes the *unique* clause resulting from repeating the following until fixpoint: if $l_1, \ldots, l_k \in C$ and there is a clause $(l_1 \vee \cdots \vee l_k \vee l) \in F \setminus \{C\}$ for some literal $l$, let $C := C \cup \{\bar{l}\}$. A clause $C \in F$ is called asymmetric blocked if ALA$(F,C)$ is blocked w.r.t. $F$. *Asymmetric blocked clause elimination* (ABCE) repeats the following until fixpoint: if there is an asymmetric blocked clause $C \in F$, let $F := F \setminus \{C\}$.

## 2 Covered Clause Elimination Procedures

Given a CNF formula $F$, a clause $C$, and a literal $l \in C$, the set of *resolution candidates* of $C$ w.r.t. $l$ is RC$(F,C,l) := \{C' \mid C' \in F_{\bar{l}}$ and $C \otimes_l C'$ is not a tautology$\}$. Notice that every clause in RC$(F,C,l)$ contains the literal $\bar{l}$. If RC$(F,C,l) = \emptyset$, then $C$ is blocked w.r.t. $F$. The literals apart from $\bar{l}$ which occur in all clauses of RC$(F,C,l)$ form the *resolution intersection* RI$(F,C,l)$ of $l$ and $C$ w.r.t. $F$, defined as

$$\text{RI}(F,C,l) := \left(\bigcap \text{RC}(F,C,l)\right) \setminus \{\bar{l}\}.$$

---
[2] Here $\bar{l} \in C$ is included to handle the special case that for any tautological *binary* clause $(l \vee \bar{l})$, both $l$ and $\bar{l}$ block the clause. Even without this addition, every *non-binary* tautological clause contains at least one literal that blocks the clause.

Given a CNF formula $F$, a clause $C \in F$, and a literal $l \in C$, we say that $l$ *covers* the literals in $\mathrm{RI}(F,C,l)$ (w.r.t. $F$ and $C$). A literal $l'$ is *covered* by $l \in C$ if $l' \in \mathrm{RI}(F,C,l)$. A literal $l \in C$ is *covering* w.r.t. $F$ and $C$ if $l$ covers at least one literal, i.e., $\mathrm{RI}(F,C,l) \neq \emptyset$.

**Lemma 1.** *For any CNF formula $F$, clause $C \in F$, and literal $l \in C$, it holds that replacing $C$ by $C \cup \mathrm{RI}(F,C,l)$ in $F$ preserves satisfiability.*

*Proof.* For any literal $l \in C$ it holds that $\mathrm{VE}(F,l) = \mathrm{VE}((F \setminus \{C\}) \cup \{C \cup \mathrm{RI}(F,C,l)\}, l)$, where $\mathrm{VE}(F,l)$ denotes the CNF formula resulting from variable eliminating[3] the variable of the literal $l$ from $F$. □

For a given clause $C$ in a CNF formula $F$, we denote by (*covered literal addition*) $\mathrm{CLA}(F,C)$ the clause resulting from repeating the following until fixpoint: if there is a literal $l \in C$ such that $\mathrm{RI}(F,C,l) \setminus C \neq \emptyset$, let $C := C \cup \mathrm{RI}(F,C,l)$.

**Lemma 2.** *Replacing a clause $C \in F$ by $\mathrm{CLA}(F,C)$ preserves satisfiability.*

*Proof.* The clause $\mathrm{CLA}(F,C)$ is obtained by iteratively applying Lemma 1 on clause $C$. □

**Lemma 3.** *Assume two clauses $C,D$ with $l \in C \subseteq D$ and two sets of clauses $F,G$ with $F \subseteq G$. Further assume that $D$ is not blocked w.r.t. $F$ and hence $C$ is not blocked w.r.t. $G$. Then $\mathrm{RC}(G,C,l) \supseteq \mathrm{RC}(F,D,l) \neq \emptyset$ and hence $\mathrm{RI}(G,C,l) \subseteq \mathrm{RI}(F,D,l)$.*

*Proof.* Monotonicity of RC w.r.t. its first argument and anti-monotonicity w.r.t. its second argument follows directly from its definition. For RI, note that intersection is anti-monotonic for non-empty sets of sets. □

**Theorem 1.** *Given a CNF formula $F$ and a clause $C \in F$, $\mathrm{CLA}(F,C)$ is blocked or uniquely defined.*

*Proof.* Assume $C$ is not blocked w.r.t. $F$ and contains two literals $l_1, l_2$, which cover the literals $L'_i = \mathrm{RI}(F,C,l_i)$ respectively. Consider the clauses $C_1 = C \cup L'_1$ and $C_2 = C \cup L'_2$. Now assume that both of $C_1, C_2$ are not blocked w.r.t. $F$. Then all clauses $D \in \mathrm{RC}(F,C_1,l_2) \subseteq \mathrm{RC}(F,C,l_2)$ contain all literals in $L'_2$. Since $C_1$ is not blocked and thus $\mathrm{RC}(F,C_1,l_2)$ is not empty, we obtain $L'_2 \subseteq \mathrm{RI}(F,C_1,l_2)$. The case where the indices are exchanged (i.e., $L'_1 \subseteq \mathrm{RI}(F,C_2,l_1)$) is symmetric. Thus as long clauses do not become blocked, covered literals can be added independently. The case that both of $C_1,C_2$ are blocked is trivial.

What remains (by symmetry) is the case that $C_2$ is blocked but $C_1$ is not. Again, as above, we get $L'_2 \subseteq \mathrm{RI}(F,C_1,l_2)$. For $C'_1 = C_1 \cup \mathrm{RI}(F,C_1,l_2)$ we have $C'_1 = C \cup L'_1 \cup \mathrm{RI}(F,C_1,l_2) \supseteq L'_1 \cup (C \cup L'_2) \supseteq C'_2$ which is also blocked. This generalizes to the following observation: For any non-deterministic choice of adding covered literals to $C$, the literal $l_2$ remains covering. Further, if in this process the clause did not become blocked, it will eventually become blocked if the covered literals of $l_2$ are added. □

To illustrate the effect of adding covered literals on logical equivalence[4], consider the formula

$$F_{\mathrm{CLA}} = (a \vee b \vee c) \wedge (a \vee \bar{b} \vee d) \wedge (a \vee \bar{c} \vee \bar{d}) \wedge (\bar{a} \vee \bar{b} \vee \bar{c}) \wedge (\bar{a} \vee b \vee \bar{d}) \wedge (\bar{a} \vee c \vee d).$$

Notice that $\mathrm{RI}(F_{\mathrm{CLA}},(a \vee b \vee c),b) = \{d\}$ and $\mathrm{RI}(F_{\mathrm{CLA}},(a \vee b \vee c),c) = \{\bar{d}\}$. Therefore, depending on the order of addition, $\mathrm{CLA}(F_{\mathrm{CLA}},(a \vee b \vee c))$ is either $(a \vee b \vee c \vee d)$ when starting with covering literal $b$ or $(a \vee b \vee c \vee \bar{d})$ when starting with covering literal $c$. In both cases $\mathrm{CLA}(F_{\mathrm{CLA}},(a \vee b \vee c))$ is blocked. After replacing $(a \vee b \vee c)$ by $(a \vee b \vee c \vee d)$ the truth assignment $\tau$ with $\tau(a) = \tau(b) = \tau(c) =$ false and $\tau(d) =$ true satisfies the new formula, while falsifying $(a \vee b \vee c) \in F_{\mathrm{CLA}}$. In fact, $F_{\mathrm{CLA}}$ witnesses the fact that none of the clause elimination procedures introduced next preserve logical equivalence in general.

---

[3] More formally, $\mathrm{VE}(F,l) = (F_l \otimes F_{\bar{l}}) \cup (F \setminus (F_l \cup F_{\bar{l}}))$, where $F_l$ and $F_{\bar{l}}$ consist of the clauses in $F$ that contain $l$ and $\bar{l}$, respectively, and $F_l \otimes F_{\bar{l}} = \{C \otimes_l C' \mid C \in F_l, C' \in F_{\bar{l}}, \text{ and } C \otimes_l C' \text{ is not a tautology}\}$.

[4] In this context, two formulas $F$ and $F'$ are logically equivalent if they have exactly the same set of satisfying assignments when restricting these assignments to the variables appearing in both $F$ and $F'$.

## 2.1 Covered Clause Elimination

**Definition 2.** *Given a CNF formula $F$, a clause $C \in F$ is* covered *if $\text{CLA}(F,C)$ is blocked w.r.t. $F$.*

**Lemma 4.** *Removal of an arbitrary covered clause preserves satisfiability.*

*Proof.* $C$ can be replaced by $\text{CLA}(F,C)$ (Lemma 2), and $C$ can be removed as $\text{CLA}(F,C)$ is blocked. □

For a given formula $F$, *covered clause elimination* (CCE) repeats the following until fixpoint: if there is a covered clause $C \in F$, remove $C$ from $F$. The resulting *unique* formula is denoted by $\text{CCE}(F)$.

Confluence of CCE follows from the following lemma.

**Lemma 5.** *The following holds for any CNF formula $F$, clause $C \in F$, and set of clauses $S \subseteq F$ such that $C \notin S$. If $C$ is covered w.r.t. $F$, then $C$ is covered w.r.t. $F \setminus S$.*

*Proof.* Let $\text{CLA}(F,C) = C_k$, where $C_0 := C$, and $C_{i+1} := C_i \cup \text{RI}(F,C_i,l_i)$ for each $i = 0..k-1$ and $l_i \in C_i$. Now define $D_0 := C$ and, for each $i = 0..k-1$, $D_{i+1} := D_i$ if $D_i$ is blocked w.r.t. $F \setminus S$ and $D_{i+1} := D_i \cup \text{RI}(F \setminus S, D_i, l_i)$ otherwise. Using Lemma 3, one can show by induction that for each $i$ we have either (i) $D_i$ is blocked w.r.t. $F \setminus S$, or (ii) $\text{RI}(F \setminus S, D_i, l_i) \supseteq \text{RI}(F,C_i,l_i)$. If (i) holds for some $i$, then $\text{CLA}(F \setminus S, C)$ is blocked w.r.t. $F \setminus C$. If $D_i$ is not blocked w.r.t. $F \setminus S$ for any $i$, then $\text{CLA}(F \setminus S, C) \supseteq \text{CLA}(F,C)$. □

**Theorem 2.** CCE *is confluent.*

**Theorem 3.** CCE *is more effective than* BCE.

*Proof.* CCE is at least as effective as BCE follows from the fact that $C \subseteq \text{CLA}(C)$: if $C$ is blocked, so is $\text{CLA}(C)$. Moreover, in $F_{\text{CLA}}$ no clause is blocked. However, all clauses are covered. Hence BCE will not remove a single clause, while CCE removes all of them. □

## 2.2 Hidden Covered Clause Elimination

For a given CNF formula $F$, a clause $C \in F$ is *hidden covered* if the clause resulting from repeating 1. $C := \text{CLA}(F,C)$; 2. $C := \text{HLA}(F,C)$ until fixpoint is blocked w.r.t. $F$. *Hidden covered clause elimination* (HCCE) repeats the following until fixpoint: if there is a hidden covered clause $C$ in $F$, remove $C$ from $F$.

**Lemma 6.** *Removal of an arbitrary hidden covered clause preserves satisfiability.*

*Proof.* Follows from the facts that (i) $F$ is satisfiability equivalent to $(F \setminus \{C\}) \cup \{\text{CLA}(F,C)\}$; (ii) $F$ is satisfiability equivalent to $(F \setminus \{C\}) \cup \{\text{HLA}(F,C)\}$; and (iii) BCE preserves satisfiability. □

**Theorem 4.** HCCE *is more effective than* CCE.

*Proof.* HCCE is at least as effective as CCE follows from the fact that $C \subseteq \text{HLA}(F,C)$: if $C$ is covered, so is $\text{HLA}(F,C)$. Moreover, consider the formula

$$F_{\text{HCCE}} = (a \vee b) \wedge (a \vee c) \wedge (\bar{a} \vee d) \wedge (\bar{a} \vee e) \wedge (b \vee c) \wedge (\bar{b} \vee d) \wedge (\bar{b} \vee \bar{e}) \wedge (\bar{c} \vee \bar{d}) \wedge (\bar{c} \vee e) \wedge (\bar{d} \vee \bar{e}).$$

In $F_{\text{HCCE}}$ no clause is covered. However, all clauses are hidden covered. Hence CCE will not remove a single clause, while HCCE removes all of them. □

By replacing CCE and BCE by HCCE and HBCE in the proof of Theorem 3 we have the following.

**Theorem 5.** HCCE *is more effective than* HBCE.

## 2.3 Asymmetric Covered Clause Elimination

For a given CNF formula $F$, a clause $C \in F$ is called *asymmetric covered* if the clause resulting from repeating 1. $C := \text{CLA}(F,C)$; 2. $C := \text{ALA}(F,C)$ until fixpoint is blocked w.r.t. $F$. *Asymmetric covered clause elimination* (ACCE) repeats the following until fixpoint: if there is a hidden covered clause $C$ in $F$, remove $C$ from $F$.

**Lemma 7.** *Removal of an arbitrary asymmetric covered clause preserves satisfiability.*

*Proof.* Follows from the facts that (i) $F$ is satisfiability equivalent to $(F \setminus \{C\}) \cup \{\text{CLA}(F,C)\}$; (ii) $F$ is satisfiability equivalent to $(F \setminus \{C\}) \cup \{\text{ALA}(F,C)\}$; and (iii) BCE preserves satisfiability. □

**Theorem 6.** ACCE *is more effective than (i)* ABCE, *and (ii)* HCCE.

*Proof.* (i) By replacing CCE and BCE by ACCE and ABCE in the proof of Theorem 3.

(ii) ACCE is at least as effective as HCCE follows from the fact that $\text{HLA}(F,C) \subseteq \text{ALA}(F,C)$: if $\text{HLA}(F,C)$ is covered, so is $\text{ALA}(F,C)$. Moreover, consider the formula

$$\begin{aligned} F_{\text{ACCE}} &= (a \vee b \vee c) \wedge (a \vee b \vee \bar{c}) \wedge (a \vee \bar{b} \vee c) \wedge (a \vee \bar{b} \vee \bar{c}) \wedge (\bar{a} \vee b \vee c) \wedge (\bar{a} \vee b \vee \bar{c}) \wedge \\ & (\bar{a} \vee \bar{b} \vee c) \wedge (\bar{a} \vee \bar{b} \vee \bar{c}) \wedge (a \vee b \vee d) \wedge (a \vee b \vee \bar{d}) \wedge (a \vee \bar{b} \vee d) \wedge (a \vee \bar{b} \vee \bar{d}) \end{aligned}$$

In $F_{\text{ACCE}}$ no clause is hidden covered. However, ACCE can remove $(a \vee b \vee c)$ and $(a \vee b \vee \bar{c})$. □

## 3 Discussion and Conclusions

Our current preliminary implementation of CCE requires on average twice the computational cost of BCE on the 2009 SAT Competition application benchmark set when applied until fixpoint. This implies that CCE can be made quite fast in practice. Regarding the practical effectiveness of CCE, on about half of the instances, $\text{CCE}(F)$ is approximately the same size as $\text{BCE}(F)$ (the difference is less than 10 clauses). However, on the other half the additional reduction is about 5% compared to BCE; for the best case, we observed one instance for which the additional reduction was as high as 40%.

As further work on this subject, we will focus on studying the effectiveness of CCE further in practice, and also possibilities of implementing HCCE and ACCE. Here it is important to notice that, even when a specific elimination technique is too costly for practical purposes to be run until fixpoint, such a technique may be of practical use in a restricted form, i.e., by only applying it on long clauses or for a restricted time. Also, we will measure the effect of applying these elimination techniques on solving interesting benchmark formulas. On the more foundational side, we will study how to reconstruct solutions for a CNF formula $F$ from solutions to any $\text{CCE}(F)$, $\text{HCCE}(F)$ and $\text{ACCE}(F)$; this is important for practical applications since CCE, HCCE, and ACCE do not preserve logical equivalence.